\documentclass[aps,prd,amsmath,amssymb,superscriptaddress,nofootinbib,onecolumn,10pt]{revtex4}
\usepackage[]{latexsym}
\usepackage[english]{babel}
\usepackage{graphicx}
\usepackage{hyperref}
\usepackage{color}
\usepackage{bm}

\def\apj{ApJ}

\def\mnras{MNRAS}
\def\physrep{PhysRep}

\def\aap{{A\&A}}
\def\prd{PRD}
\def\jcap{JCAP}

\newcommand{\ca}{}

\newcommand{\h}{\mathcal{H}}
\newcommand{\so}{\sigma_0}
\newcommand{\To}{\tau_0}
\newcommand{\mI}{\mathcal{I}}
\newcommand{\mJ}{\mathcal{J}}
\newcommand{\mQ}{\mathcal{Q}}

\newcommand{\Dk}[1]{\frac{d^3#1}{(2\pi)^3}} 
\newcommand{\dk}[1]{\delta^{(1)}_0(\ve #1)} 

\newcommand{\ve}[1]{{\text{\bf #1}}} 
\newcommand{\vk}{\ve k}
\newcommand{\vp}{\ve p}
\newcommand{\vq}{\ve q}
\newcommand{\vx}{\ve x}

\begin{document}

\title{The dark matter dispersion tensor in perturbation theory}
\author{Alejandro Aviles}
\email{avilescervantes@gmail.com}
\affiliation{Department of Astronomy, University of California, Berkeley, CA 94720, USA.}

\begin{abstract}
We compute the dark matter velocity dispersion tensor up to third order in perturbation theory using the Lagrangian formalism,
revealing growing solutions at the third and higher orders.
Our results are general and can be used for any other perturbative formalism.
As an application, corrections to the matter power spectrum are calculated, \ca and we find that some of them have
the same structure as those in the effective field theory of large-scale structure, with ``EFT-like'' coefficients that
\ca grow quadratically with the linear growth function and are further suppressed by powers of the 
logarithmic linear growth factor $f$; other corrections present  
additional $k$ \ca dependence. Due to the velocity dispersions, there exists a free-streaming scale that suppresses the whole 1-loop power spectrum.
Furthermore, we find that as a consequence of the nonlinear evolution, the free-streaming length is shifted towards larger scales, wiping out more structure
than that expected in linear theory.
Therefore, we argue that the formalism
developed here is better suited for a perturbation treatment of warm dark matter or neutrino clustering, 
where the velocity dispersion effects are well known to be important.
We discuss implications related to the nature of dark matter. 
\end{abstract}

\maketitle

\section{Introduction}

The large-scale clustering of dark matter is well understood at the early stages of the Universe's history, when matter perturbations are small
and linear theory is reliable. As the Universe evolves, matter perturbations grow by gravitational instability and the 
scales of interest---probed by recent and current observations---reach the onset of the nonlinear regime. 
Several models beyond linear theory have been developed since time ago in order to gain physical insight and proper interpretation of the outcomes 
of these observations and of N-body simulations; see, e.g., Refs.~\cite{Sahni:1995rm,Ber02,Cooray:2002dia}.
One of these methods is Perturbation Theory (PT), in which the central idea is to formally expand matter and velocity fields and to 
generate solutions to the field equations order by order. 
These fields are generally proportional to the $n$-{\it \'esime} power of the linear fields, which grow with time.
Also, matter density fluctuations have a scale dependence such that small structures become nonlinear earlier.
Thus, the full expansion loses its validity when higher orders become larger and larger 
and PT becomes meaningless; 
for a review on PT see \cite{Ber02}. Nevertheless, it is still interesting to study the ``quasilinear''  
regime within the framework of PT: here, the theory is
reliable and the different versions, especially Standard Perturbation Theory (SPT)
 \cite{Juszkiewicz01121981,Vishniac01061983,Fry:1983cj,Goroff:1986ep,Jain:1993jh,Carlson:2009it}, Lagrangian Perturbation Theory (LPT) 
\cite{Zel70,Buc89,Mou91,Catelan:1994ze,BouColHivJus95,Catelan:1996hw,TayHam96,Bha96,Mat08a,Mat08b,Okamura:2011nu,Carlson:2012bu,TasZal12a,Rampf:2012xa,Whi14,Vlah:2014nta},
and several others purposed to obtain better convergence and performance 
\cite{Crocce:2005xy,Crocce:2005xz,McDonald:2006hf,Matarrese:2007wc,Pietroni:2008jx,Bernardeau:2008fa,Anselmi:2010fs,Bernardeau:2011dp,Anselmi:2012cn,Blas:2013aba}, 
lead to notable improvements
to the linear theory at large scales. Closer to the nonlinear scales, the situation changes drastically, and it is 
not clear at which order we start to neglect higher orders \cite{Carlson:2009it,Vlah:2014nta}, making the theory unreliable.

The main problem of PT is that in nonlinear theory all Fourier modes are coupled and even those
out of its reach can affect the large scales, which can be seen as integrations over all 
momenta in loop calculations. To 
tame this problem some schemes beyond PT have been proposed. 
One notable case is the effective field theory of large-scale structure
(EFTofLSS), both in the Eulerian \cite{BNSZ12,CHS12,Hertzberg:2012qn,Pajer:2013jj,Carrasco:2013mua,CLP14,Foreman:2015lca,Baldauf:2015aha} 
and Lagrangian \cite{Porto:2013qua,McQuinn:2015tva, Vlah:2015sea, SenZal15,Baldauf:2015tla,Baldauf:2015zga,Zaldarriaga:2015jrj}
pictures. These theories are valid only at large scales, but by {\it integrating out} the ultraviolet (UV) physics, not amenable by PT, 
they incorporate backreaction effects upon long wavelength modes. 
As a result, the dark matter is provided with an effective stress tensor, leading to a nonperfect fluid description with 
viscosity and pressure perturbations. 
In the Lagrangian formalism one can physically visualize the EFT as a theory of a smoothed collection
of dark matter particles with effective internal structure, sourcing the Poisson equation with a multipolar field expansion; 
as a result, long wavelength perturbations experience tidal forces from the UV parametrized physics.
The underlying idea of EFTofLSS is simple and powerful, and it has shown improvements over the
SPT and LPT, at the cost of introducing free parameters that should be estimated by comparing to observations or to N-body simulations.
Other approaches, akin to EFTofLSS, consider from the very beginning effective imperfect viscous fluids, arriving 
at similar results \cite{Floerchinger:2014jsa,Blas:2015tla,Fuhrer:2015cia}.

During the first stages of gravitational collapse, the dark matter is single streaming; therefore, a perfect fluid model is 
well motivated and the stress tensor can be safely neglected. Later, nonlinear collapse makes different 
streams to converge, leading to larger velocity dispersions, the formation of matter caustics and ultimately to {\it shell crossing}. 
At shell crossing LPT breaks down, and SPT probably does, too: 
LPT and SPT have been probed to be equivalent up to third order for the matter power spectrum \cite{Mat08a} and the matter bispectrum \cite{Rampf:2012xa}, 
and in one dimension to all orders \cite{McQuinn:2015tva}.

In this paper we propose a route to continue the development of PT by truncating 
the Boltzmann hierarchy at the second moment and by introducing
the velocity dispersion tensor (VDT) into the field equations. 
Nonlinear velocity dispersions for cold dark matter (CDM) have been considered previously, 
but either with the use of phenomenological effective descriptions or by considering only the scalar mode, the trace, of the VDT
\cite{Buchert:2005xj,Pueblas:2008uv,Shoji:2009gg,McDonald:2009hs}. 
In contrast, in this paper we consider the full 2-rank VDT derived directly from the Boltzmann equation of collisionless, classical particles. 
It is expected that the convergence of the field expansion series 
suffers similar convergence problems as in SPT and LPT, 
but we show that new corrections to the matter power spectrum arise that can be important, especially in the cases of warm dark matter (WDM) and 
neutrinos. We compute the VDT
up to third order in PT using the Lagrangian formalism, finding growing solutions starting at third order. 
Our results depend only on the linear matter density fields and therefore can be applied to any other perturbation scheme. 
The introduction of the VDT leads to 25 corrections to the Lagrangian displacement field; all of them are implicitly written, some of them
are explicitly computed, and their corresponding corrections to the matter power spectrum are calculated. We find that some of these corrections have the same form
as those in the 1-loop EFTofLSS, scaling as $-\ell^2(\tau) k^2 P_L$, with $\ell^2$ depending on time as $D_+^2$, and further suppressed by powers 
of the logarithmic linear growth factor $f \equiv d \ln \delta^{(1)}/ d \ln a$. Other
corrections with additional $k$ dependence are found.

All our results depend on a small dimensionless quantity: $\so$, the trace of the zero-order VDT evaluated today. A 
{\it free-streaming} scale $\lambda_\text{fs} \sim \sqrt{\so}/\h_0$ is introduced, below which dark matter clustering is suppressed. 
If the dark matter is a thermal relic, $\so$ can, in principle,
be related to the temperature to mass ratio at kinetic decoupling. Nevertheless, being a very small quantity, it is susceptible to 
non-negligible corrections from higher orders in PT, higher momenta in the Boltzmann hierarchy we are neglecting, as well as 
backreaction effects upon its background value \cite{Floerchinger:2014jsa,McDonald:2009hs}. In this sense, $\so$ 
could be interpreted as a free parameter as much as those in the EFTofLSS.
The linear theory of CDM velocity dispersions has been studied recently in \cite{Piattella:2015nda}, suggesting lower bounds
of about $\so \sim 10^{-14}$ for thermally produced dark matter. Nevertheless, this value seems large compared to some particle CDM theories where this 
value could be as small as $10^{-25}$ \cite{Bringmann:2006mu,Piattella:2015nda}. If the latter is the case, the formalism developed here gives negligible corrections to PT. 
The case of WDM is different \cite{Bode:2000gq,Colin:2000dn,Hansen:2001zv,Zavala:2009ms,Schneider:2013ria}; 
here, the dark matter particles are still relativistic at the freezing-out epoch, and velocity dispersions are 
still important at the stages of the formation of galaxies, inducing considerably large free-streaming scales.
One effect is the appearance of an effective free-streaming scale with $\lambda_\text{eff\,fs} > \lambda_\text{fs}$. This is a consequence
of the nonlinear evolution of the VDT, which does not decay simply as $a^{-2}$, but instead presents growing modes that suppress the power spectrum
at somewhat larger scales than in the linear theory.

We argue that our formalism can be applied to neutrinos' clustering during their nonrelativistic stages. 
In this case $\so$ (usually called $\sigma^2_\nu$ in neutrino literature) is larger, and it is well known to play an important 
role suppressing the total matter clustering in both linear \cite{Bond:1980ha, Hu:1997mj,Crotty:2004gm,Lesgourgues:2006nd,Shoji:2010hm}
and nonlinear \cite{Saito:2008bp,Wong:2008ws,Saito:2009ah,Lesgourgues:2009am,Bird:2011rb,AliHaimoud:2012vj,VillaescusaNavarro:2012ag,Blas:2014hya,Peloso:2015jua}
theories. In an appendix to this paper we estimate the free-streaming scale of neutrinos within our formalism, 
finding good agreement with standard estimations in the literature; see e.g. \cite{Ringwald:2004np,Shoji:2010hm}.

The scales considered throughout this paper are well below the Hubble horizon, while General Relativity corrections are of the order
$k/aH$ and therefore safely within Newtonian theory.
Furthermore, in calculations we assume an Einstein-de Sitter (EdS) universe both for the sake of 
simplicity and because the main stages of structure formation take place
during the matter-dominated era. Nevertheless, we keep track of the $f$ and $D_+$ functions to obtain a 
better approximation beyond EdS---this procedure has been proven to give accurate results in PT.

The rest of the paper is organized as follows. 
In Sec.~\ref{section:BE} the Boltzmann hierarchy is revised; in Sec.~\ref{section:EoM} we write the field
equations in the Lagrangian picture; in Sec.~\ref{section:VDT} we outline the calculations for the VDT; in Sec.~\ref{section:LD} 
we obtain the corrections to the Lagrangian displacement field and to the matter power spectrum; 
in Sec.~\ref{section:conclusions} we present our conclusions and future perspectives. 
We delegate the main calculations to appendixes.

\section{Boltzmann Equation}  \label{section:BE}

We start by considering the dark matter as a collection of $n$ collisionless, classical, nonrelativistic particles, each
with position $\vx_n(\tau)$ and conjugate momentum $\vp_n(\tau)$. The phase-space distribution function is
\begin{equation} \label{BE:f}
 f({\bf x}, {\bf p }, \tau) = \sum_n \delta_D (\ve x- \ve x_n(\tau)) \delta_D (\ve p - \ve p_n(\tau)),
\end{equation} 
or some smoothed version of it,
such that $f$ is the number of particles in the phase-space volume $d^3x d^3p$. Its evolution is dictated by the 
collisionless Boltzmann equation
\begin{equation} \label{BE:BE}
 \frac{df}{d\tau}({\bf x}, {\bf p }, \tau) = \frac{\partial f}{\partial \tau} +  \frac{1}{a m} \ve p \cdot \frac{\partial f}{\partial \ve x} -
 a m \frac{\partial \phi}{\partial \ve x} \cdot \frac{\partial f}{\partial \ve p} = 0,
\end{equation}
and by the Poisson equation
\begin{equation} \label{BE:PE}
 \partial^2 \phi(\ve x, \tau) = \frac{3}{2} \mathcal{H}^2 \Omega_m \delta.
\end{equation}

In the fluid approximation we take moments of the phase-space distribution to get the density 
\begin{equation} \label{BE:rho}
 \rho(\ve x, \tau) = m a^{-3} \int d^3 p f 
\end{equation}
and the local mean peculiar velocity
\begin{equation}  \label{BE:meanvel}
 \rho \langle u^i \rangle_p (\ve x, \tau) = a^{-4} \int d^3 p p^i f. 
\end{equation}
From now on, $v^i \equiv  \langle u^i \rangle_p $. The notation $\langle \,\cdot\, \rangle_p = \int d^3 p \,(\,\cdot\,) f / \int d^3 p f $
denotes momentum average over the ensemble.
The second momentum of the distribution provides the stress tensor
\begin{equation}  \label{BE:stressTdef}
\rho  \langle u^i u^j \rangle_p = a^{-5} \int d^3 p \frac{1}{m} p^i p^j f  = \rho v^i v^j + \rho \sigma^{ij}.
\end{equation}
In the last equality, we decomposed the stress tensor with the introduction of the VDT 
\begin{equation} \label{BE:stressT}
\sigma^{ij} = \langle \Delta u^i \Delta u^j \rangle_p,
\end{equation}
with $\Delta u^i(\ve x, \tau) \equiv   v^i - u^i $. Similarly, one can construct higher rank tensors
\begin{equation} \label{BE:vddef}
\sigma^{ijk} \equiv \langle \Delta u^i \Delta u^j \Delta u^k \rangle_p = - \langle u^i u^j u^k\rangle_p + v^{\{i} \sigma^{jk\}} + v^iv^jv^k,
\end{equation}
where $T^{\{\cdots\}}$ indicates sum over cyclic permutations of indices, and
\begin{equation}  \label{BE:3moment}
 \rho  \langle u^i u^j u^k \rangle_p = a^{-6} \int d^3 p \frac{1}{m^2} p^i p^j p^kf. 
\end{equation}

The evolution equations follow from taking moments of Boltzmann equation (\ref{BE:BE}) and using Eqs.~(\ref{BE:rho})--(\ref{BE:3moment}), 
obtaining the continuity and Euler equations
\begin{equation}  \label{BE:continuity}
 \partial_\tau \delta(\ve x, \tau)  + \partial_j ((1+\delta)v^j) = 0,
\end{equation}
\begin{equation} \label{BE:euler}
 \partial_\tau v^i(\ve x, \tau)  + v^j \partial_j v^i + \mathcal{H}v^i + \partial^i \phi = -\frac{1}{\rho}\partial_j (\rho \sigma^{ij}),
\end{equation}
and the VDT equation
\begin{equation}  \label{BE:vdeq}
 \partial_\tau \sigma^{ij}(\ve x, \tau)  + 2 \mathcal{H} \sigma^{ij} + v^{k} \partial_k \sigma^{ij} + \sigma^{ik}\partial_k v^j
 + \sigma^{jk}\partial_k v^i = \frac{1}{\rho}\partial_k (\rho \sigma^{ijk}).
\end{equation}
We can generate an infinite hierarchy of equations in this way,
each one coupled to its previous and posterior momenta. In this work we close the system of equations 
by setting the rank-3 dispersion to zero: $\sigma^{ijk} =0$.
From Eq.~(\ref{BE:vdeq}) we can observe that for small initial perturbations the VDT decays as $\propto a^{-2}$; moreover, if initially the VDT  
vanishes then it will remain zero during the whole evolution of the above equations. 
In Sec.~\ref{section:VDT} we shall see that as nonlinear 
evolution becomes important, the VDT starts to grow. 

\section{Lagrangian description of the Equations of motion}  \label{section:EoM}

In a Lagrangian description \cite{Zel70} we follow the trajectories $\ve x$ of dark matter particles with position $\ve q$ at some initial time $\tau_i$,  such that
\begin{equation} \label{lagtoeul}
 \ve x (\ve q, \tau) = \ve q + \ve s(\ve q, \tau), \qquad \qquad   \ve s(\ve q, \tau_i) =0,
\end{equation}
where $\ve s$ is the Lagrangian map. The peculiar velocity of each particle is
\begin{equation}
 \ve u (\ve x , \tau) = \frac{d \ve x}{d\tau} = \frac{d \ve s (\ve q, t)}{d\tau}.  
\end{equation}
We define the Lagrangian displacement field $\Psi^i(\ve q, \tau)$ through its derivative as
\begin{equation}
 \dot{\Psi}^i(\ve q, \tau) \equiv \langle \dot{\ve s}^i(\ve q, \tau) \rangle_p =  v^i.
\end{equation}

If there were no velocity dispersions, then the $\Psi(\ve q,\tau)$ and $\ve s(\ve q,\tau)$ fields would have the same functional dependence. The residual allows us
to write $\sigma^{ij}$ as a function of Lagrangian coordinates as

\begin{equation}
 \sigma^{ij}(\ve q, \tau) = \langle (\dot{\Psi} - \dot{\ve s})^i(\dot{\Psi} - \dot{\ve s})^j\rangle_p.
\end{equation}

The difference between the two vectors is a {\it stochastic residual} vector field
\begin{equation}
 \Gamma^i(\ve q, \tau) = \ve s^i - \Psi^i. 
\end{equation}
In LPT literature it is assumed that $\Gamma^i =0$, but the emergence of a nonzero value is unavoidable when velocity dispersions are included.
Now, using Eqs.~(\ref{BE:euler}) and (\ref{BE:vdeq}) and changing from partial to total time derivatives we get the equations of motion, 
\begin{equation} \label{LPT1:EOM}
 \ddot{\Psi}^i (\ve q, t) + \mathcal{H} \dot{\Psi}^i + \partial^i \phi (\ve q + \Psi) =-\frac{1}{1+\delta}\partial_j ((1+\delta) \sigma^{ij}),
\end{equation}

\begin{equation}  \label{LPT1:stressEq}
 \dot{\sigma}^{ij} (\ve q, t)   + 2 \mathcal{H} \sigma^{ij}  + \sigma^{ik}\partial_k \dot{\Psi}^j
 + \sigma^{jk}\partial_k \dot{\Psi}^i = 0.
\end{equation}
Throughout this paper the symbols $\partial$ and $\nabla$ denote derivatives with respect to Eulerian and Lagrangian coordinates respectively, 
and $J^i_{\,j}$ is the Jacobian matrix
of the transformation between them, such that $\nabla_i = J^{k}_{\,\,i}\partial_i.$ We also thoroughly use the relation 
\begin{equation}\label{BE:deltaPsi}
 1+\delta(\vx) = \frac{1}{\text{det}( I + \nabla \Psi(\vq) )},
\end{equation}
between the Eulerian density contrast and the Lagrangian displacement. To arrive at this result we use the
mass conservation equation $(1+\delta(\vx)) d^3 x = (1+\delta(\vq)) d^3 q$, and assume a relation
 $\bm{J} = \bar{\bm{J}} ( 1- \delta(\vq))$
between the determinants of the Jacobian matrices 
\begin{equation}\label{JacMat}
 J^i_{\,j} = \delta^{i}_{j} + \nabla_j s^i \qquad \text{and} \qquad \bar{J}^i_{\,j} \equiv  \delta^{i}_{j} + \nabla_j \Psi^i.
\end{equation}
As usual, we can choose the initial time $\tau_i$ such that $\delta(\vq)$ is negligible. Thus, the condition 
$\bm{J} = \bar{\bm{J}}$ implies that the contribution of $\Gamma^i$ to the coordinate transformation (\ref{lagtoeul}) is to locally rotate the
system. We further discuss the role of the stochastic residual field in Appendix \ref{app::Gamma}.

In LPT, the right-hand side (rhs) of Eq.~(\ref{LPT1:EOM}) vanishes and the solution is $\Psi^{i} = \sum \Psi^{(n) i}$ 
\cite{Catelan:1994ze,BouColHivJus95,Catelan:1996hw}, 
with\footnote{Throughout this paper we adopt the shorthand notation  
\begin{equation}
 \int_{\vk}  = \int \prod_{i=1}^{n} \frac{d^3 k_i}{(2\pi)^3} (2\pi)^3 \delta_D(\ve k - \vk_1 - \cdots - \vk_n).  
\end{equation}} 
\begin{equation} \label{LPT1:LDFS}
 \Psi^{(n) i}(\vk, \tau) = i\frac{D_+^n}{n!}\int_{\vk} L^{(n) i}(\vk_1, \dots, \vk_n) \delta^{(1)}_0(\vk_1) \cdots \delta^{(1)}_0(\vk_n),
\end{equation}
where $D_+(\tau,\tau_0)$ is the linear growth function, such that $\delta^{(1)}(\vk,\tau) = D_+(\tau,\tau_0) \delta^{(1)}_0(\vk)$.
Recursive relations for the $L^{(n)}$ kernels are calculated in \cite{Rampf:2012up,Matsubara:2015ipa}, 
and up to order $n=3$, they are written below in Eq.~(\ref{LPTkernels}). The first-order solution gives the Zel'dovich
approximation \cite{Zel70}.

\section{Approximations for the velocity dispersion tensor} \label{section:VDT} 

The approach we follow in the perturbative analysis is to iteratively solve Eq.~(\ref{LPT1:stressEq}) 
up to third order by using the well-known solutions  $\Psi^{(1)}$,
$\Psi^{(2)}$, and  $\Psi^{(3)}$ for the displacement field in LPT. Strictly, we should have to solve both 
Eqs.~(\ref{LPT1:EOM}) and (\ref{LPT1:stressEq}) simultaneously order by order, but
we appeal to the fact that while the Lagrangian displacements are growing functions of time at all orders, the velocity dispersion tensor
is not. More importantly, the VDT background solution introduces a small dimensionless constant, 
$\so$, which further suppresses the rhs of Eq.~(\ref{LPT1:EOM}), and
therefore all the terms we are neglecting in our approach are quadratic and cubic in $\so$, as we see below.
In Sec.~\ref{section:LD}, once we have the solutions to the VDT,  we find some of the induced corrections to the Lagrangian displacement and to the matter power
spectrum.

As it is standard in PT, we formally expand the VDT as
\begin{equation}
 \sigma^{ij} = \sigma^{(0)ij} + \sigma^{(1)ij} + \sigma^{(2)ij} +\sigma^{(3)ij} + \cdots,
\end{equation}
and we search for solutions to the equation 
\begin{equation}  \label{AVD:stressEq}
 \dot{\sigma}^{(n)ij}  + 2 \mathcal{H} \sigma^{(n)ij} =  S^{(n)ij} \equiv
 - \Big[\sigma^{ik} [J^{-1}]^{l}_k \nabla_l \dot{\Psi}^{j}\Big]^{(n)} + \,\, (\,\,i \leftrightarrow j \,\,)\,\,
\end{equation} 
with an iterative procedure.
The zero-order solution (with $S^{(0)ij}=0$) is, by isotropy and homogeneity of the background,
\begin{equation} \label{VDT::S0order}
 \sigma^{(0)ij} \propto \delta^{ij} a^{-2} \propto D_+^{-2} \propto \h^4, 
\end{equation}
where the last two proportionalities are valid in an EdS universe. 
The next step is to plug this solution into the order $n=1$ equation (\ref{AVD:stressEq}). Given that $\dot{\Psi}^{(1)} = f \h \Psi \propto \h D_+$,
the source at order $n=1$ goes as $\h^5 D_+ \propto \h^3$, and solving Eq.~(\ref{AVD:stressEq}) we obtain 
$\sigma^{(1)} \propto f \h^2 \propto D_+^{-1} $. This process is iterated to subsequent orders, each getting an additional $D_+(\tau,\To)$ time dependence.

In Appendix \ref{app::solvesigma} all the sources $S^{(n)ij}$ and the solutions $\sigma^{(n)ij}$ are calculated up to order $n=3$. We 
write here the latter in Fourier space\footnote{Our convention of the Fourier transform is 
\begin{equation} \label{FourierSpaceConv}
 f(\ve x) = \int \frac{d^3 k}{(2 \pi)^3} f(\ve k) e^{i\ve k \cdot \ve x} \qquad \Rightarrow \qquad 
 (2\pi)^3 \delta_D^3 (\ve x - \ve x') = \int d^3 k\, e^{ i \ve k \cdot (\ve x - \ve x') }.
\end{equation}
}
as
\begin{eqnarray} \label{sol0123orders}
 \sigma^{(0)}_{ij}(\vk, \tau)&=& \frac{1}{3} \delta_{ij} \so\left(\frac{a_0}{a(\tau)}\right)^2 \delta_D(\vk), \label{sol0orders}\\
 & & \nonumber\\
 \sigma^{(1) ij}(\vk, \tau) &=& \frac{2}{3} \so \left(\frac{a_0}{a(\tau)}\right)^2 f D_+(\tau,\To)  
  \frac{k^i k^j}{k^2} \delta^{(1)}_0(\ve k),  \label{sol1orders}\\
 & & \nonumber\\
\sigma^{(2)ij}(\vk, \tau) &=& \frac{1}{3} \so\left(\frac{a_0}{a(\tau)}\right)^2 f D_+^2(\tau,\To) \mI^{ij}(\ve k), \label{sol2orders}\\
 & & \nonumber   \\
 \sigma^{(3)ij}(\vk, \tau) &=& \frac{1}{9}\so \left(\frac{a_0}{a(\tau)}\right)^2 f D_+^3(\tau,\To) \mJ^{ij}(\ve k). \label{sol3orders}
\end{eqnarray}
It is good to keep track of the $f$, $D_+$ and scale factors in the solutions, because they arise from different aspects: $a^{-2}$ from
the zero-order solution, the $D_+$'s from each displacement field, and the $f$'s from the derivatives of the displacement fields.
The background and first-order solutions, Eqs.~(\ref{sol0orders}) and (\ref{sol1orders}) respectively, were found previously in \cite{McDonald:2009hs} using SPT.

 The matrices $\mI$ and $\mJ$ are
\begin{eqnarray} 
\mI^{ij} (\ve k) &=&  \frac{1}{2} \int_{\vk}   
K^{(2)ij}(\ve k_1, \ve k_2) \dk{k_1} \dk{k_2}, \\
 & & \nonumber\\
\mJ^{ij}(\ve k) &=& \frac{1}{2}  \int_{\ve k} 
 K^{(3)ij}(\ve k_1, \ve k_2, \ve k_3)  \dk{k_1} \dk{k_2} \dk{k_3}, 
\end{eqnarray}
with the kernels $K^{(2)ij}$ and $K^{(3)ij}$ given by
\begin{eqnarray} \label{K2andK3}
  K^{(2)ij}(\ve k_1, \ve k_2) &=& (1+2f) (\vk_1 \cdot \vk_2) L^{(1) i}(\vk_1)  L^{(1)j }(\vk_2) 
  +  k^{i} L^{(2) j}(\ve k_1, \ve k_2) + \,\, (\,\,i \leftrightarrow j \,\,)\,\,, \label{K2andK3K2} \\
 & & \nonumber\\
K^{(3)ij}(\ve k_1, \ve k_2, \ve k_3) &=& 2(1+f)(1+2f) \frac{(\ve k_1 \cdot \ve k_2)(\ve k_2 \cdot \ve k_3)}{k^2_2}L^{(1) i}(\vk_1)  L^{(1)j }(\vk_3) \nonumber\\
  &+& 3(1+2f) \ve k_1 \cdot (\ve k_2 + \ve k_3) \,L^{(1) i }(\vk_1) L^{(2)j } (\ve k_2, \ve k_3) 
  +  k^{i } L^{(3)j} (\ve k_1, \ve k_2, \ve k_3)  +  \,\, (\,\,i \leftrightarrow j \,\,)\,\,, \label{K2andK3K3}
\end{eqnarray}  
and $L^{(1)}$, $L^{(2)}$ and $L^{(3)}$ are the standard kernels in LPT 
\cite{Catelan:1994ze,BouColHivJus95,Catelan:1996hw,Rampf:2012up,Matsubara:2015ipa}: 

\begin{eqnarray} \label{LPTkernels}
L^{(1)}(\vk) &=&  \frac{\vk}{k^2}, \qquad \qquad
L^{(2)}(\vk_1,\vk_2) =  \frac{3}{7}\frac{\vk}{k^2}\left[1- \left(\frac{\vk_1 \cdot \vk_2}{k_1 k_2} \right)^2 \right],
    \label{LPTL12}\\
 & & \nonumber\\ 
L^{(3)}(\vk_1,\vk_2,\vk_3) &=&  \frac{5}{7}\frac{\vk}{k^2}\left[1- \left(\frac{\vk_1 \cdot \vk_2}{k_1 k_2} \right)^2 \right]
\left[1- \left(\frac{(\vk_1+\vk_2) \cdot \vk_3}{|\vk_1 + \vk_2| k_3} \right)^2 \right]\label{LPTL3} \\
& & -\frac{1}{3}\frac{\vk}{k^2} \left[1 - 3 \left(\frac{\vk_1 \cdot \vk_2}{k_1 k_2} \right)^2 
+ 2 \frac{(\vk_1\cdot\vk_2)(\vk_2\cdot\vk_3)(\vk_3\cdot\vk_1)}{k_1^2 k_2^2 k_3^2}\right], \nonumber
\end{eqnarray}
with $\vk = \vk_1 + \cdots + \vk_n$. In the expression for $L^{(3)}$ there is an additional transverse term that we omit because it is not
needed at the lowest order. For calculation purposes, it is convenient to symmetrize the $K^{ij}$ kernels.

Equations (\ref{sol0orders})--(\ref{sol3orders}) are the first main results of this paper.
We have found the VDT up to third order, revealing growing solutions starting at third order. 
To obtain the VDT we have used LPT. 
Nevertheless, note that the results are independent of the PT formalism because they are 
expressed purely in terms of the linear density contrast; 
they are independent in the same sense that the relations of the Lagrangian displacement 
field and the density contrast, written in Eq.~(\ref{LPT1:LDFS}), are. 
In the rest of the paper we explore some consequences of this result.

\section{Corrections to the Lagrangian displacement field}  \label{section:LD} 

In this section we find the corrections to the Lagrangian displacement field using the method of Green's functions.
Since we work only with longitudinal modes, it is convenient to define, using the notation of \cite{Porto:2013qua},
\begin{equation} \label{deftheta}
 \theta(\vq ,\tau) \equiv \nabla_i \Psi^i(\vq, \tau). 
\end{equation}
By taking the divergence with respect to the Lagrangian coordinates to Eq.~(\ref{LPT1:EOM}), we get the equation of motion for $\theta$,
\begin{equation} \label{LPT1:EOMt}
 \ddot{\theta} (\ve q, t) + \mathcal{H} \dot{\theta} - \frac{3}{2}\Omega_m \h^2\theta = \mathcal{Q}_{\text{LPT}} + \mathcal{Q},
 \end{equation}
where the standard nonlinear source in LPT is $\mathcal{Q}_{\text{LPT}}  = -J^{ij} \partial_j\partial_i \phi(\vq + \Psi) - \frac{3}{2}\Omega_m \h^2\theta$, 
and the sources of the ``new'' terms are $\mathcal{Q} = \sum \mathcal{Q}^{(n)}$ with
\begin{eqnarray} 
 \mathcal{Q}^{(n)}(\ve q,\tau) &=& -\left[ \nabla_i \Big( \det(I +  \nabla\cdot\Psi) [J^{-1}]^k_j \nabla_k 
 (\det(I +  \nabla\cdot\Psi)^{-1} \sigma^{ij}) \Big) \right]^{(n)} \nonumber\\
 &=& -\nabla_i  \sum_{p,q,r,s} \det(I +  \nabla\cdot\Psi)^{(p)} [J^{-1(q)}]^k_j \nabla_k 
 (\det(I +  \nabla\cdot\Psi)^{-1(r)} \sigma^{(s)ij})  \label{DefQSources}  \\
 &\equiv& \sum_{p,q,r,s} \mathcal{Q}^{(pqrs)}(\ve q,\tau), \label{DefQSources2} 
\end{eqnarray}
where the orders are constrained by $n = p+q+r+s$. Noting that $r+s \ge 1$, we get that at order $n=3$ 
there are 16 different terms in the sum; at $n=2$, 7 terms;
and at $n=1$, 2 terms.
The time dependence of $\mathcal{Q}^{(n)}$ is proportional to $D_+^p \times D_+^q \times D_+^r \times  D_+^{-2+s}$ modulo the $f$ factors,
or
\begin{equation} \label{sQTD}
\mathcal{Q}^{(n)} \propto D_+^{n-2}.
\end{equation}

The solutions to Eq.~(\ref{LPT1:EOMt}) can be formally written as
\begin{equation} \label{exptheta}
 \theta = \theta^{(1)} + \theta^{(2)} + \theta^{(3)}  + \cdots + \theta_{\mQ^{(1)}} + \theta_{\mQ^{(2)}} + \theta_{\mQ^{(3)}}  + \cdots, 
\end{equation}
where $\theta^{(i)}$ are the standard solutions in LPT and
\begin{equation} \label{thetaQn}
 \theta_{\mQ^{(n)}} \equiv \sum_{p,q,r,s} \theta^{(pqrs)}.
\end{equation}
To find the $\theta_{\mathcal{Q}^{(n)}_i}$ functions,
we integrate their respective sources in Eq.~(\ref{DefQSources2}) against the advanced 
Green's function of the linear operator defined by the left-hand side of Eq.~(\ref{LPT1:EOMt}), 
\begin{equation} \label{Gfunction}
 G(a_0,a) = -\frac{2}{5}\frac{1}{\h^2_0 a_0}\left[\frac{a}{a_0} - \left( \frac{a}{a_0}\right)^{-3/2} \right] \Theta_H(a_0-a).
\end{equation}
By considering only the fastest growing solution, it follows that
\begin{equation} \label{Gfunctionsol}
 \int_{a}^{a_0} G(a_0,a') \left(\frac{a'}{a_0}\right)^m d a'= \frac{2 }{5(2+m)\h^2_0 }\left(\frac{a}{a_0}\right)^{m+2}. 
\end{equation}
Using Eqs.~(\ref{sQTD}) and (\ref{Gfunctionsol}) we obtain
\begin{equation}
 \theta_{\mathcal{Q}^{(n)}_i} = \frac{2}{5 n } \mathcal{Q}^{(n)}_i \frac{D_+^2}{\h^2_0} \propto D_+^{n}.
\end{equation}
That is, the corrections have the same time dependence as the LPT original solutions times additional $f$ factors---the 
exceptions are terms with $s=0$ in their sources, which do not suffer $f$ suppression.

In Appendix \ref{app::correlations} we find the spectrum (\ref{P_1_0003}) 
(this is the only contribution affected by the third-order VDT),
which we rewrite here as
\begin{eqnarray} \label{contQ3::1_1wwwww}
 \langle \theta^{(1)}(\ve k,\tau)  \theta^{(0003)}(\ve k',\tau) \rangle' &=& 
 - \frac{2}{675}f \left(\frac{94}{7} +\frac{26}{3} f(3+2f) \right) \sigma^{2}_{L}(\tau) \frac{ \so}{\h_0^2} 
  k^2 P_L(\vk, \tau)              \\
 &-& \frac{2}{135}f  \left(\frac{k^3}{4\pi^2} \int_0^\infty dr r^2 P_L(kr,\tau) \tilde{R}_{(0003)}(r) \right)
 f \frac{\so}{\h^2_0} k^2 P_L(\vk, \tau) , \nonumber
\end{eqnarray}
where  the function $\tilde{R}_{(0003)}(r)$ is defined in Eq.~(\ref{defineR}) and plotted in Fig.~\ref{fig:PlotsR}. 
Here, $\sigma^{2}_{L}$ is the total linear matter density fluctuations variance,
\begin{equation}
 \sigma^{2}_{L}(\tau) = \int \frac{d^3k}{(2\pi)^3} P_L(k,\tau).
\end{equation}
This integral is UV divergent for $P_L \propto k^n$ with $n\ge - 3$. 
This is the case for the usual linear matter power spectra, where $n \approx -2.1$ at small scales. 
Nevertheless, the power spectrum is expected to have a cutoff at high $k$ due to dark matter velocity dispersion 
(see, e.g., Ref.~\cite{Piattella:2015nda}), 
making the integral convergent. The effect is analogous to the free-streaming cutoff into the neutrino linear power spectrum.

Therefore, we have found one of several modifications to the linear PS:

\begin{equation} \label{ps1corr}
 P_{nl}(\ve k, \tau) = P_L(\ve k, \tau) + \cdots \,\, - \ell_{v.d.}^2(\tau)  k^2 P_L(\ve k, \tau) 
 -A(k, \tau)  k^2 P_L(\ve k, \tau)\,\,+ \,\,\cdots
\end{equation}
where the time dependence is
\begin{equation}
 A(k, \tau), \,\,\ell_{v.d.}^2(\tau) \propto  D_+^2(\tau)
\end{equation}

In EFTofLSS the modifications are quite similar: the EFT coefficients are negative and proportional to $k^2 P_L(\ve k,\tau)$. Moreover, 
the time dependence is very similar. For example, by using scaling arguments on self-similar solutions 
in Ref.~\cite{Pajer:2013jj} (see also Sec.~3.2 of Ref.~\cite{Baldauf:2015tla}), 
a $D^2_{+}$ time dependence is expected for $P_{L} \propto k^{-2}$. The same 
conclusion is obtained in \cite{Porto:2013qua} by renormalization arguments. Also, there are claims that $\ell_{EFT}^2$ should
scale as its dimensions do, as $[\,\text{length}\,]^2$. Additionally, we have found a correction, $A(k, \tau)$, that has a further $k$ dependence.

All the corrections can be computed as outlined above,
and all of them have the same time dependence (modulo the $f$ factor) ---with 
the exception of the terms arising from the spectrum $\langle\theta^{(1)} \theta_{\mathcal{Q}_i^{(1)}} \rangle$, 
scaling as $D_+^2$.

Using Eqs.~(\ref{DefQSources}) and (\ref{Gfunctionsol}) we obtain the linear correction
\begin{equation} \label{lcto1}
 \theta_{\mathcal{Q}_1}(\vk,\tau) = \theta^{(0010)} + \theta^{(0001)} = \frac{2}{5} \frac{1+2f}{3}\frac{\so}{\h^2_0} k^2 \dk{\vk} D_+(\tau)
 = -\frac{2}{5} \frac{1+2f}{3}  \frac{\so}{\h^2_0} k^2\theta^{(1)}(\vk,\tau).
\end{equation}
Accordingly, the linear displacement field is modified as
\begin{equation} \label{lcdf}
 \Psi^{(1)i} \rightarrow \Psi^{(1)i} + \Psi^{(1)i}_{\mathcal{Q}_1} =  \Psi^{(1)i} -\frac{2}{5}\frac{1+2f}{3}\frac{\so}{\h^2_0}k^2\Psi^{(1)i}
\end{equation}
where we used $\theta_{\mathcal{Q}_1} = i k_i\Psi^{(1)i}_{\mathcal{Q}_1}$. 

Now, we are in the position to estimate the terms we are neglecting in our perturbative approach. If we plug the linear displacement field 
correction in Eq.~(\ref{lcdf}) into the source [given by Eq.~(\ref{app:sources2order})] of the VDT equation at second order, the induced corrections to
the VDT second-order solution would acquire powers of $\so$ greater than 1. This shows that all the terms we are neglecting
by sourcing the VDT equation with the well-known LPT solution are small, as we anticipated in Sec.~\ref{section:VDT}.

The linear correction term [Eq.~(\ref{lcto1})] also gives rise to
\begin{equation} \label{P13Corr}
  \langle \theta^{(3)}(\ve k,\tau) \theta_{\mathcal{Q}_1}(\ve k',\tau) \rangle' = 
  -\frac{2}{5}\frac{1+2f}{3}  \frac{\so}{\h^2_0} k^2 k^i k^j C^{(31)}_{ij}(\vk) D_+^4(\tau)
\end{equation}
where the function $C^{(13)}(\vk)$ is defined in terms of the spectra of the displacement 
field in Matsubara's resummation scheme \cite{Mat08a} [in general, the
polyspectra are defined in Eq.~(\ref{MatPol})].
Similarly, we can pair $\theta^{(2)}(\ve k,\tau)$ with the second-order terms  
that contain the $L^{(2)}$ LPT  kernel, obtaining
\begin{equation} \label{P22Corr1}
 2 \langle \theta^{(2)}(\ve k,\tau) (\theta^{(0002)}(\ve k',\tau) +\theta^{(0020)}(\ve k',\tau) )\rangle' 
  \ni -\frac{2}{5}\frac{1+2f}{3}  \frac{\so}{\h^2_0} k^2 k^i k^j C^{(22)}_{ij}(\vk) D_+^4(\tau)
\end{equation}
where the rhs includes only the $L^{(2)}$ piece of the second-order VDT in $\mathcal{Q}^{(0002)}$ and the second-order Lagrangian displacement field in 
$\mathcal{Q}^{(0020)}$.
From Eqs.~(\ref{lcto1}) and (\ref{MatPol}) we calculate the corrections to the bispectrum of the displacement field,
\begin{equation} \label{P211Corr1}
 C^{(211)}_{ijk}(\vk_1,\vk_2,\vk_3) \rightarrow C^{(\bar{2}\bar{1}\bar{1})}_{ijk}(\vk_1,\vk_2,\vk_3) 
 \ni \left(1- \frac{2}{5}\frac{1+2f}{3}  \frac{\so}{\h^2_0} k^2 \right) C^{(211)}_{ijk}(\vk_1,\vk_2,\vk_3),
\end{equation}
(here we are following the definition of barred orders given in Appendix \ref{app::correlations}).
Therefore we expect a suppression of the whole 1-loop power spectrum by the free-streaming scale $k_{\text{fs}}$, 
defined in Eq.~(\ref{FSscale}) of Appendix \ref{app:fss}.

\begin{figure}
	\begin{center}
	\includegraphics[width=4 in]{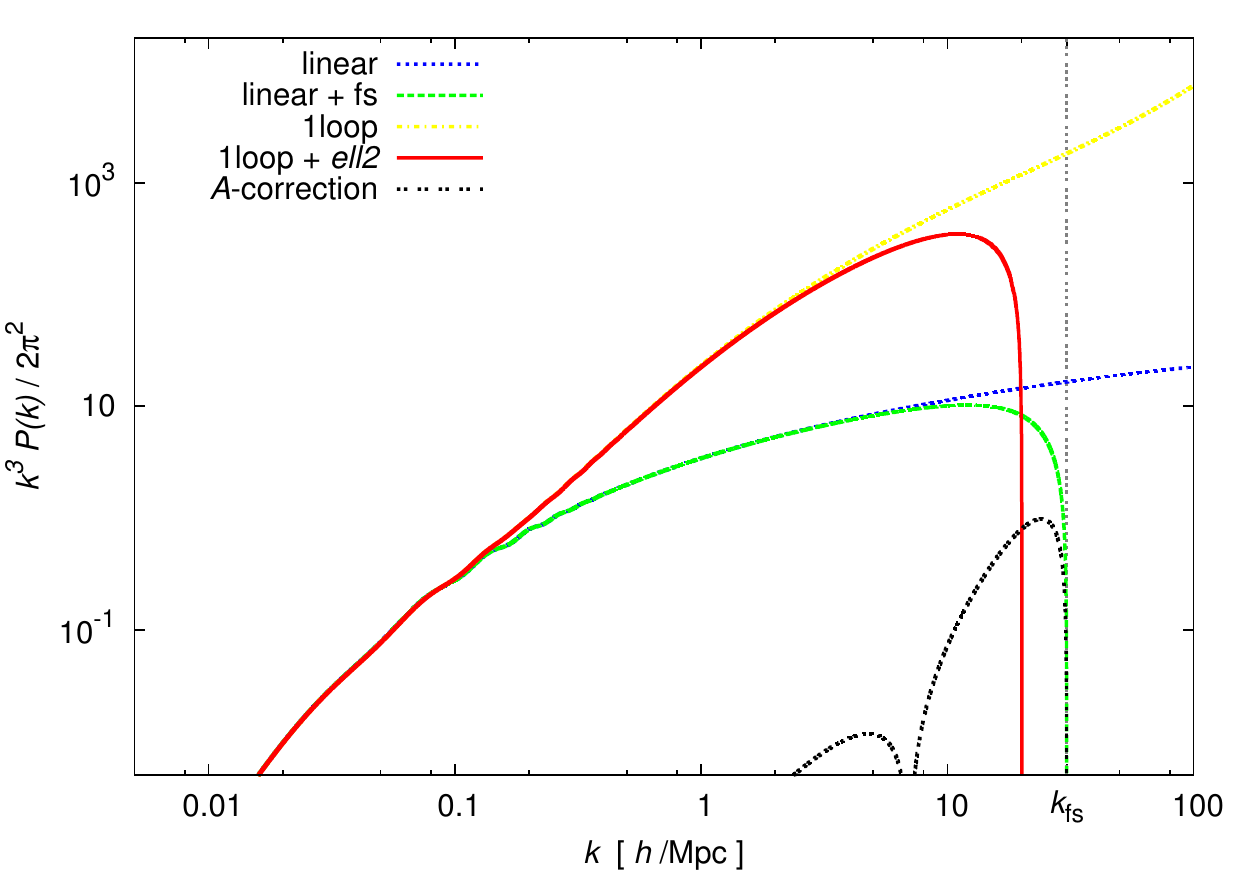}
	\caption{Power spectrum corrections in Eq.~(\ref{ps1corr}). 
	Here we show $(1-\ell^2_{v.d.} k^2)  P_L(k) + P_{\text{1-loop}}(k)$ (solid red),
	the linear power spectrum $P_L(k)$ (dotted blue line), the linear correction 
	given by Eq.~(\ref{pslc}) (dashed green line), the 1-loop power spectrum 
	(dot-dashed yellow line), and $|A(k)|k^2P_L(k)$ (dashed black line). 
	The vertical line shows the free-streaming scale. We note the appearance of
	an effective free-streaming scale at a larger scale to that given by the linear theory. See text for details. 
	\label{fig:PS}}
	\end{center}
\end{figure}

Corrections such as those given by Eqs.~(\ref{P13Corr}), (\ref{P22Corr1}) and (\ref{P211Corr1}) may not considerably affect the power spectrum (\ref{ps1corr}), 
or they are at least subdominant. This is because they are not enhanced by the matter linear fluctuations 
variance $\sigma^{2}_{L}$, as it happens with the correction $-\ell_{v.d.}^2 k^2 P_L$ in Eq.~(\ref{ps1corr}).
To show this, we consider an example with $\so = 10^{-10}$, corresponding to a 
free-streaming scale $k_{\text{fs}} = 30.5 \, h/\text{Mpc}$ and 
$\ell^2_{v.d.} = 0.2$ at redshift $z=0$ (this 
could be, for example, the case of WDM gravitinos with small masses $\sim 1\,\text{keV}$ 
\cite{Zavala:2009ms,Takahashi:2007tz}). 
We compute the corrections in Eq.~(\ref{ps1corr}) by replacing the linear power spectrum 
by its correction given in Eq.~(\ref{pslc});
we checked that this gives almost identical numerical results as cutting off the integrals at the
free-streaming scale. The linear power spectrum is obtained from the code CAMB \cite{Lewis:1999bs} using the Planck 2015 best-fit 
cosmological parameters \cite{Ade:2015xua}. 
The results are shown in Fig.~\ref{fig:PS}: 
the solid red line corresponds to $(1-\ell_{v.d.}^2 k^2)P_L(k) + P_{\text{1-loop}}$, 
the dotted blue line corresponds to the linear power spectrum 
$P_L(k)$, the dashed green line
is the linear power spectrum corrected by Eq.~(\ref{ps1corr}), the dot-dashed yellow line is the 1-loop power spectrum, and the
dashed black line is $|A(\vk)|k^2 P_L(k)$. The vertical dashed line shows the free-streaming scale as given by Eq.~(\ref{FSscale}).
We note that the leading correction is given by the $\ell_{v.d.}^2$ term; 
this happens because here the free-streaming
length scale squared ($1/k_{\text{fs}}^2$) is enhanced by the total variance $\sigma^{2}_L$, which does not occur in the $A(k)$ correction.
The same argument holds for the 
corrections in Eqs.~(\ref{P13Corr}), (\ref{P22Corr1}) and (\ref{P211Corr1}). They are suppressed only by
the free-streaming scale, and their corrections to the power spectrum will 
be subdominant---hence we only plot the 1-loop without its cutoff. 
We note again from Fig.~\ref{fig:PS} that the free-streaming scale is corrected to an effective one at a 
larger scale; physically this is a consequence of the nonlinear evolution of the VDT, which is growing at third order and, hence, wiping out more structure.
Nevertheless, to study this effect in full quantitative detail we have to compute all the corrections to the power spectrum 
in Eq.~(\ref{MatPS}). We leave such analysis to future investigations.


\section{Conclusions} \label{section:conclusions} 

In this work we have continued the development of PT by calculating the contributions of the 
VDT which sources Euler's equation. Our perturbative approach considers that the standard LPT
solutions for the Lagrangian displacement $\Psi^{(1)}$, $\Psi^{(2)}$ and $\Psi^{(3)}$ 
are not affected in a first approximation, and we plug them into the VDT 
equation to solve it iteratively order by order. After calculating the VDT up to the 
third order we use them as sources of the Lagrangian displacement field
equation and obtain the corrections to the LPT solutions. We argue that this approach leads to a good approximation 
for two reasons: first, while the
matter Lagrangian displacement field grows with time at all orders, the VDT does not; second, the VDT is suppressed by a small dimensionless constant $\so$, 
introduced since the background solution, and therefore all the corrections we are neglecting are quadratic or cubic in $\so$.

We find that the VDT has growing solutions beginning at the third order. The solutions have the form
of convolutions of the linear matter density fields, and their time dependence scales as $D_+^{2-n}$ 
at order $n$ with $D_+$ the linear growth function; they are further suppressed by
powers of the logarithmic linear growth factor $f$. Although calculated  in LPT, the solutions 
can be used as input for any other perturbative scheme. Moreover, the generality of the VDT solutions allows us to use them for 
studying neutrinos clustering in their nonrelativistic evolution, where the effects of the velocity dispersion are even more important, 
broadening the domain of applicability of our formalism.

The VDT corrects the displacement field with 25 ``new'' terms, all of them implicitly written in this work. 
16 are third order, 7 second order, and 2 first order.
We explicitly calculate a few of them and find the induced corrections to the power spectrum. 
The results of these corrections are quite similar 
to those in the 1-loop EFTofLSS, with the 
correct $k^2$ dependence, and grow as the EFT coefficients do. We also find that other corrections provide additional $k$ dependence.
Furthermore, the whole 1-loop power spectrum in LPT (or SPT), without the VDT, is suppressed by a free-streaming scale when the velocity dispersions are 
considered.
To find all the corrections to the power spectrum, 
it is necessary to calculate the 25 sources, integrate them against the Green's function and take the correlations
among all of them. We delegate the detailed calculations of all corrections to a future work. 

The leading corrections to the matter power spectrum are multiplied by the small constant $\so$ 
but also by the matter density linear fluctuations variance $\sigma^2_{L}$. 
The latter is divergent if velocity dispersions are not considered, but the linear power spectrum 
has a cutoff at high $k$ when they are present. 
One of the consequences of these terms is a shift of the free-streaming scale, 
suppressing the formation of structure at larger scales,
as we have seen in this paper for WDM with velocity dispersion 
$\so = 10^{-10}$, although a precise estimation of this nonlinear effect
requires the full power spectrum in the presence of velocity dispersions. 

Regarding the case of cold dark matter, if we assume it is a thermal relic, before kinetic decoupling its temperature decays
like that of baryons in the early Universe, that is, as 
$\propto a^{-1}$, and after that it decays as $\propto a^{-2}$; moreover, the value of $\so$ can be connected to the temperature to mass ratio 
of dark matter particles at 
kinetic decoupling. Therefore, an estimation of $\so$ would provide extremely useful information about the fundamental nature of dark matter and its
interactions with the standard model of particles in the early Universe. This seems a formidable task---if not 
impossible---if the  dark matter is cold and therefore $\so$
very small such that baryon nongravitational effects would mask its influence.

One alternative is that of WDM, in which considerably lighter particles can
suppress the halo formation just below the scales of dwarf galaxies \cite{Schneider:2013ria}. Other dark matter candidates that could
provide a large free-streaming scale are, for example, WDM-CDM mixtures \cite{delaMacorra:2009yb,Maccio':2012uh} and 
scalar field Bose-Einstein condensates \cite{Hu:2000ke,Suarez:2013iw,Alcubierre:2015ipa}.

Nowadays there exists a vivid discussion regarding several aspects of WDM N-body simulations; 
see, for example, Refs.~\cite{Lovell:2013ola,Paduroiu:2015jfa} 
and reference therein. One of these concerns is about initial conditions, which are usually given by the linear (cuttedoff) power spectrum, 
bulk velocities given by 2LPT \cite{Crocce:2006ve}, and zero (or underestimated) velocity dispersions; 
the latter because they are smaller than bulk Zel'dovich flows. Recently, it was argued in 
\cite{Paduroiu:2015jfa} that this approach 
is incorrect, and if velocity dispersions are considered appropriately in the initial conditions, 
the structure formation qualitatively 
changes and an interplay between a standard hierarchical top-down model at large scales 
and a bottom-up model at small scales is observed. 
We further argue that our predicted effects cannot be fully observed in simulations with vanishing (or underestimated) initial velocity dispersions. 
Although, we note that nonlinear features, especially those generated after shell crossing, 
cannot be captured by our formalism but can be observed in simulations.

The case of neutrinos is special because their linear velocity dispersion is considerably larger, $\sigma_{\nu,i 0} \sim 10^{-7} \text{eV}^2 / m_{\nu,i}^2$, and
an accurate measurement of it, represents one of the most promising methods to infer the neutrinos' absolute masses \cite{Hu:1997mj,Beutler:2014yhv}, in contrast
to their squared mass  differences obtained from solar and atmospheric oscillation experiments.

\begin{acknowledgments}
The author would like to thank to Zvonimir Vlah, Yu Feng and Martin White for discussions 
and useful suggestions. A.~A.~ is supported by the
UCMEXUS-CONACYT Postdoctoral Fellowship.
\end{acknowledgments}

\appendix

\section{CALCULATION OF THE SOURCES AND SOLUTION OF THE VELOCITY DISPERSION TENSOR} \label{app::solvesigma}

In this appendix we perturbatively solve for the VDT, $\sigma^{ij}$. 
From Eq.~(\ref{AVD:stressEq}) we write the source of the VDT equation at order $n$ as 
\begin{equation} \label{app:sources}
 S^{(n)ij}(\ve q,\tau)  = [-\sigma^{ik}\partial_{k}\dot{\Psi}^j]^{(n)} \,\, + \,\, (i \leftrightarrow j )
 =   - \sum_{p,q,r} \sigma^{(p)ik} [J^{-1 (q)}]^{l}_k \nabla_l \dot{\Psi}^{(r)j} + \,\, (\,\,i \leftrightarrow j \,\,)\,\,.
\end{equation}
Here, the orders $p$, $q$ and $r$ are related through $n =p+q+r$. 
Since the order of $\dot{\Psi}$ is at least $r=1$, the order of $\sigma^{ij}$ 
should be at most $p = n -1$.
$J$ is the Jacobian matrix of the coordinate transformation $\ve x= \ve x(\ve q)$, given in Eq.~(\ref{JacMat}).
Expanding the sources up to the first three orders we get
\begin{eqnarray}
 S^{(1) ij} &=& [-\sigma^{ik}\partial_{k}\dot{\Psi}^{j}]^{(1)} \, + \,\, (i \leftrightarrow j )  =
 -\sigma^{(0)ik}\nabla_{k}\dot{\Psi}^{(1)j} + \,\, [\,\,\text{s.t}\,\,]^{(1)} \, + \,\, (i \leftrightarrow j ) \label{app:sources1order} \\
 &\equiv& \, S^{(1) ij}_A +  S^{(1) ij}_\Gamma \,,  \nonumber\\
 & & \nonumber\\
 S^{(2) ij} &=& [-\sigma^{ik}\partial_{k}\dot{\Psi}^{j}]^{(2)}  \,\, + \,\, (i \leftrightarrow j )  \label{app:sources2order}\\
 &=& [-\sigma^{ik}\nabla_{k}\dot{\Psi}^{j} + \sigma^{ik}\nabla_{k}\Psi^{l} \nabla_{l} \dot{\Psi}^{j} ]^{(2)} 
 + [\,\,\text{s.t}\,\,]^{(2)} \,\, + \,\, (i \leftrightarrow j ) \nonumber\\
 &=& -\sigma^{(0)ik}\nabla_{k}\dot{\Psi}^{(2)j}  -\sigma^{(1)ik}\nabla_{k}\dot{\Psi}^{(1)j}
  + \sigma^{(0) ik}\nabla_{k}\Psi^{(1) l} \nabla_{l} \dot{\Psi}^{(1) j} 
   + \,\,\,\, [\,\,\text{s.t}\,\,]^{(2)} \,\, + \,\, (i \leftrightarrow j ) \nonumber\\
 &\equiv& \, S^{(2) ij}_A + S^{(2) ij}_B + S^{(2) ij}_C + S^{(2) ij}_\Gamma \,, \nonumber\\
  & & \nonumber\\
 S^{(3) ij} &=& [-\sigma^{ik}\partial_{k}\dot{\Psi}^{j}]^{(3)}  \,\, + \,\, (i \leftrightarrow j ) \label{app:sources3order}\\
 &=&  [-\sigma^{ik}\nabla_{k}\dot{\Psi}^{j} + \sigma^{ik}\nabla_{k}\Psi^{l} \nabla_{l} \dot{\Psi}^{j}  
 -  \sigma^{ik}\nabla_{k}\Psi^{l} \nabla_{l} \Psi^{s}\nabla_{s} \dot{\Psi}^{j} ]^{(2)} + \,\,\,\, [\,\,\text{s.t}\,\,]^{(3)} \,\, \,\, + \,\, (i \leftrightarrow j ) \nonumber\\
 &=& -\sigma^{(0)ik}\nabla_{k}\dot{\Psi}^{(3)j}  -\sigma^{(1)ik}\nabla_{k}\dot{\Psi}^{(2)j} -\sigma^{(2)ik}\nabla_{k}\dot{\Psi}^{(1)j} \nonumber\\
 & & + \sigma^{(0) ik}\nabla_{k}\Psi^{(1) l} \nabla_{l} \dot{\Psi}^{(2) j} + \sigma^{(0) ik}\nabla_{k}\Psi^{(2) l} \nabla_{l} \dot{\Psi}^{(1) j}
  + \sigma^{(1) ik}\nabla_{k}\Psi^{(1) l} \nabla_{l} \dot{\Psi}^{(1) j} \nonumber\\
 & &  -  \sigma^{(0)ik}\nabla_{k}\Psi^{(1)l} \nabla_{l} \Psi^{(1)s}\nabla_{s} \dot{\Psi}^{(1)j}  
 + \,\,\,\, [\,\,\text{s.t}\,\,]^{(3)} \,\, + \,\, (i \leftrightarrow j ) \nonumber\\
 &\equiv& \, S^{(3) ij}_A + S^{(3) ij}_B + S^{(3) ij}_C +S^{(3) ij}_D + S^{(3) ij}_E + S^{(3) ij}_F + S^{(3) ij}_G
 + S^{(3) ij}_\Gamma \,,\nonumber
\end{eqnarray}
where $S^{(n)ij}_\Gamma$ are the sources due to the $\Gamma$ stochastic terms ($\,[\,\,\text{s.t}\,\,]\,$)---not 
explicitly written here, but they arise from the change from Eulerian to Lagrangian derivatives. 
We will not consider these terms in this appendix; we delegate this discussion to 
Appendix \ref{app::Gamma}, where we show that their inclusion leads to subdominant corrections.

To integrate the VDT equation we assume an EdS universe; thus, Eq.~(\ref{AVD:stressEq}) takes the form 
\begin{equation}  \label{AppB:diffeq}
 \dot{\sigma}(\tau) + \frac{4}{\tau} \sigma = \frac{C}{\tau^m} \qquad \Rightarrow \qquad  \sigma = \frac{C}{(5-m)\tau^{m-1}} + \frac{c_I}{\tau^4},
\end{equation}
where $C$ is the time-independent factor of the source and $c_I$ is an integration constant that we drop since this leads to the solution of the
background equation. 

For the derivatives of the Lagrangian displacement field we extensively use the relation
\begin{equation} \label{app:LDFder}
 \dot{\Psi}^{(n)i} = n f \h \Psi^{(n)i},
\end{equation}
which follows from Eq.~(\ref{LPT1:LDFS}).

From the two previous equations and the discussion after Eq.~(\ref{VDT::S0order}), we infer an structure of sources and solutions of the form

\begin{eqnarray}
 S^{(0)} = 0 \qquad & \rightarrow & \qquad
  \sigma^{(0)} = c^{(0)}  \so \tau_0^4 \h^4  \propto  D_+^{-2} \\
  S^{(1)} \propto \h^5 D_+ \propto \frac{1}{\tau^3}  \qquad & \rightarrow & \qquad
  \sigma^{(1)} = c^{(1)} \so f \tau_0^2 \h^2 [\delta^{(1)}_0] \propto  f D_+^{-1} \\
  S^{(2)} \propto \h^5 D_+^2 \propto \frac{1}{\tau}  \qquad & \rightarrow & \qquad
 \sigma^{(2)} = c^{(2)}  \so f [\delta^{(1)}_0 *\delta^{(1)}_0 ] \propto  f \mathcal{P}_1(f) \\
  S^{(3)} \propto \h^5 D_+^3 \propto \tau  \qquad & \rightarrow & \qquad
 \sigma^{(3)} = c^{(3)} \so f \frac{1}{\tau_0^2 \h^2} [[\delta^{(1)}_0 *\delta^{(1)}_0 ] *\delta^{(1)}_0 ]  \propto  
    f \mathcal{P}_2(f) D_+.
\end{eqnarray}
The schematically depicted convolutions arise from changing to Fourier space, and they will be clarified below.
We can also note that the sources at order $n$ are proportional to a polynomial of $f$ of degree $n$. Each $f$ appears
when taking the derivative of the displacement field, by repeated use of Eq.~(\ref{app:LDFder}).

\subsection{Zero order}

In this case the source of the VDT equation vanishes. Considering the isotropy and homogeneity of the background expansion, 
the solution to Eq.~(\ref{LPT1:stressEq}) is

\begin{equation} \label{A1:sol0order}
 \sigma^{(0)}_{ij}(\tau) = \frac{1}{3} \delta_{ij} \so\left(\frac{a_0}{a(\tau)}\right)^2
 =  \frac{1}{48} \delta_{ij} \so\tau^4_0 \mathcal{H}^4.
\end{equation}
where the last equality is valid in an EdS universe.

\subsection{First order}

From Eqs.~(\ref{app:sources1order}) and (\ref{A1:sol0order}), the source in Fourier space is given by

\begin{equation}
 S^{(1) ij} = \frac{1}{6}\so f \tau_0^2 \h^3 \frac{k^i k^j}{k^2}  = \frac{C_{1}(\ve k)}{\tau^3}, 
 \qquad C_1 = \frac{4}{3}\so f \tau_0^2  \frac{k^i k^j}{k^2}.
\end{equation}
We integrate the VDT equation by using Eq.~(\ref{AppB:diffeq}) to obtain
\begin{equation} \label{A::sol1order}
 \sigma^{(1) ij}(\ve k, \tau) = \frac{1}{6} \so f \tau_0^2  \mathcal{H}^2 \delta^{(1)}_0(\ve k)
  \frac{k^i k^j}{k^2}.
\end{equation}

\subsection{Second order}

First, we calculate all the sources in Fourier space.
Using Eq.~(\ref{app:sources2order}) they are

\begin{equation}
 S^{(2)ij}_I(\ve k) =  \frac{2}{3}\so f  \h \mI^{ij}_{I},    \qquad \qquad I \in \{A,B,C \}
\end{equation}
with the $ \mI^{ij}_I$ matrices
\begin{eqnarray}
\mI^{ij}_{A} (\ve k) &=& \frac{1}{2} \int_{\vk}  
(k^i L^{(2) j} (\ve k_1, \ve k_2) + k^j L^{(2) i} (\ve k_1, \ve k_2))\dk{k_1} \dk{k_2}, \\
\mI^{ij}_{B} (\ve k) &=& \frac{1}{2}\cdot 2 f \int_{\vk} 
\frac{\ve k_1 \cdot \ve k_2}{k_1^2 k_2^2}(k^i_1 k^j_2 +k^j_1 k^i_2 )\dk{k_1} \dk{k_2}, \\
\mI^{ij}_{C} (\ve k) &=& \frac{1}{2}\int_{\vk} 
\frac{\ve k_1 \cdot \ve k_2}{k_1^2 k_2^2}(k^i_1 k^j_2 +k^j_1 k^i_2 )\dk{k_1} \dk{k_2}.
\end{eqnarray}

By noting the symmetries of the sources, we sum the three contributions to obtain

\begin{equation}
 S^{(2)ij}(\ve k) = \frac{2}{3}\so f  \h \mI^{ij},
\end{equation}
with
\begin{equation} 
\mI^{ij} (\ve k)=  \frac{1}{2} \int_{\vk} 
K^{(2)ij}(\ve k_1, \ve k_2) \dk{k_1} \dk{k_2},
\end{equation}
and the kernel $K^{(2)ij}$ is\footnote{Note that $K^{ij}(\ve k_1,\ve k_2) = K^{ji}(\ve k_1,\ve k_2) = K^{ij}(\ve k_2,\ve k_1)$ 
(the last equality holding inside the integral). Also $K^{ij}(\ve k,\ve k) = K^{ij}(\ve k,-\ve k) = (1+2f)  k^i  k^j /k^2$.}
\begin{eqnarray}
 K^{(2)ij}(\ve k_1, \ve k_2) &=& (1+2f) \frac{\ve k_1 \cdot \ve k_2}{k_1^2 k_2^2}(k^i_1 k^j_2 + k^j_1 k^i_2)  +    
(k^i L^{(2) j} (\ve k_1, \ve k_2) + k^j L^{(2) i} (\ve k_1, \ve k_2)). 
\end{eqnarray}
Now we integrate the VDT equation using the result in Eq.~(\ref{AppB:diffeq}) to obtain the second-order VDT:

\begin{equation} \label{AppB:2orderSol}
\sigma^{(2)ij}(\ve k, \tau) = \frac{1}{3} \so f  \mI^{ij}(\ve k),  
\end{equation}
with its only time dependence given by the logarithmic growth factor $f$.

\subsection{Third order}

Using Eq.~(\ref{app:sources3order})  we compute the sources at third order. These are given by
\begin{equation}
 S^{(3)ij}_I (\ve k)=  \frac{4}{3}\frac{\so f}{\To^2  \h} \mJ^{ij}_I,    \qquad \qquad I \in \{A,B,C,D,E,F,G \},
\end{equation}
with the $ \mJ^{ij}_I$ matrices

\begin{eqnarray}
  \mJ_A^{ij}(\ve k) &=& \frac{1}{2}  \int_{\vk} 
 k^{\{i } L^{(3)j\}} (\ve k_1, \ve k_2, \ve k_3)  \dk{k_1} \dk{k_2} \dk{k_3} \\
\mJ_B^{ij}(\ve k) &=& \frac{1}{2} \cdot 4 f \int_{\vk} 
\frac{\ve k_1 \cdot (\ve k_2 + \ve k_3)}{k_1^2} \,k_1^{\{i } L^{(2)j\} } (\ve k_2, \ve k_3) \dk{k_1} \dk{k_2} \dk{k_3} \\
\mJ_C^{ij}(\ve k) &=& \frac{1}{2}\cdot f \int_{\vk} 
\frac{k_{1 k} }{k_1^2} k_1^{\{ i}  K^{(1) j\}k }(\ve k_2, \ve k_3 ) \dk{k_1} \dk{k_2} \dk{k_3} \\
\mJ_D^{ij}(\ve k) &=& \frac{1}{2} \cdot 2  \int_{\vk} 
\frac{\ve k_1 \cdot (\ve k_2 + \ve k_3)}{k_1^2} \,k_1^{\{i } L^{(2)j\} } (\ve k_2, \ve k_3) \dk{k_1} \dk{k_2} \dk{k_3} \\
\mJ_E^{ij}(\ve k) &=& \frac{1}{2}  \int_{\vk} 
 \frac{\ve k_1 \cdot (\ve k_2 + \ve k_3)}{k_1^2} \,k_1^{\{i } L^{(2)j\} } (\ve k_2, \ve k_3)  \dk{k_1} \dk{k_2} \dk{k_3} \\
\mJ_F^{ij}(\ve k) &=& \frac{1}{2} \cdot 4 f  \int_{\vk} 
 \frac{(\ve k_1 \cdot \ve k_2)(\ve k_2 \cdot \ve k_3)}{k^2_1k^2_2k^2_3} k^{\{i }_1 k^{ j \}  }_3  \dk{k_1} \dk{k_2} \dk{k_3} \\
 \mJ_G^{ij}(\ve k) &=& \frac{1}{2} \cdot 2  \int_{\vk} 
 \frac{(\ve k_1 \cdot \ve k_2)(\ve k_2 \cdot \ve k_3)}{k^2_1k^2_2k^2_3} k^{\{i }_1 k^{ j \}  }_3  \dk{k_1} \dk{k_2} \dk{k_3} 
\end{eqnarray}

By noting that we can split $\mJ_C$ into different contributions as
\begin{eqnarray}
\mJ_C^{ij}(\ve k) &=& \frac{1}{2}\cdot 2 f(1+2f) \int_{\vk}
\frac{(\ve k_1 \cdot \ve k_2)(\ve k_2 \cdot \ve k_3)}{k^2_1k^2_2k^2_3} k^{\{i }_1 k^{ j \}  }_3
\dk{k_1} \dk{k_2} \dk{k_3} \\
&+&  \frac{1}{2}\cdot 2 f \int_{\vk} 
 \frac{\ve k_1 \cdot (\ve k_2 + \ve k_3)}{k_1^2} \,k_1^{\{i } L^{(2)j \} } (\ve k_2, \ve k_3)
 \dk{k_1} \dk{k_2} \dk{k_3},\nonumber
\end{eqnarray}
we sum all contributions to obtain
\begin{equation} \label{app:Jij}
\mJ^{ij}(\ve k) = \frac{1}{2}  \int_{\vk} 
 K^{(2)ij}(\ve k_1, \ve k_2, \ve k_3)  \dk{k_1} \dk{k_2} \dk{k_3}, 
\end{equation}
with
\begin{eqnarray}\label{app:KK3}
  K^{(3)ij}(\ve k_1, \ve k_2, \ve k_3) &=&  k^{\{i } L^{(3)j\}} (\ve k_1, \ve k_2, \ve k_3)
  + 3(1+2f)\frac{\ve k_1 \cdot (\ve k_2 + \ve k_3)}{k_1^2} \,k_1^{\{i } L^{(2)j\} } (\ve k_2, \ve k_3) \nonumber\\
  &+& 2(1+f)(1+2f) \frac{(\ve k_1 \cdot \ve k_2)(\ve k_2 \cdot \ve k_3)}{k^2_1k^2_2k^2_3} k^{\{i }_1 k^{ j \}  }_3 
\end{eqnarray}
And the VDT solution to  third order is
\begin{equation}\label{sigma3Solution}
 \sigma^{(3)ij}(\ve k,\tau) = \frac{4}{9} \frac{\so f}{\To^2 \h^2} \mJ^{ij}(\ve k), 
\end{equation}
which is the first order that present a growing mode.

\section{CORRECTIONS TO THE MATTER POWER SPECTRUM} \label{app::correlations}

We first give a brief review of the calculation of the matter power spectrum in LPT; we refer the reader to
Refs.~\cite{TayHam96,Mat08a,Carlson:2012bu,Vlah:2014nta} for details.
In LPT the matter power spectrum is written in terms of the Lagrangian displacement field as \cite{TayHam96}
\begin{equation}\label{app2:PLPT}
 P^{\text{LPT}}(\vk) = \int d^3 q e^{-i\vk \cdot \vq} \left( \langle e^{-i\vk \cdot \Delta}\rangle - 1 \right),
\end{equation}
where $\Delta^i = \Psi^i(\vq_2) - \Psi^i (\vq_1)$, and $\vq = \vq_2-\vq_1$.
We rewrite Eq.~(\ref{app2:PLPT}) using the cumulant expansion theorem and considering up to third-order 
displacement fields,

\begin{equation}\label{app2:PLPT1loop}
 (2\pi)^3\delta_D(\vk) +  P^{\text{LPT}}(\vk) = \int d^3 q  e^{-i\vk \cdot \vq}  \exp \left[- \frac{1}{2} k_i k_j \langle \Delta_i \Delta_j \rangle_c 
  + \frac{i}{6}k_i k_j k _k \langle \Delta_i \Delta_j \Delta_k \rangle_c \right].
\end{equation}

The way we expand the exponential in Eq.~(\ref{app2:PLPT1loop}) leads to different resummation schemes. 
In CLPT \cite{Carlson:2012bu}, for example, the terms that have zero limits as $q \rightarrow \infty $ 
are expanded; such a procedure keeps the 
cumulants $\langle \Delta^i\Delta^j \rangle_c$ in the exponential and expands the rest.

We note that the products  $\Delta^i\Delta^j$ have contributions at the same point, or zero lag, and contributions at a separation $\vq$. 
In iPT \cite{Mat08a}, the zero-lag contributions are kept in the exponential, arriving at 
\begin{eqnarray} \label{MatPS}
 P_\text{1-loop}^{\text{LPT}}(\vk) &=& \text{exp}\left[ -k_i k_j \int \frac{d^3p}{(2\pi)^3} C^{(11)}_{ij}(\vp) \right] \nonumber \\
   & \times&    \Bigg[ k_i k_j \big( C^{(11)}_{ij}(\vk) + C^{(22)}_{ij}(\vk) + C^{(31)}_{ij}(\vk) + C^{(13)}_{ij}(\vk) \big)  \nonumber\\
   &+ &    k_ik_jk_k  \int \frac{d^3p}{(2\pi)^3}\big( C^{(112)}_{ijk}(\vk, -\vp, \vp-\vk) + 
 C^{(121)}_{ijk}(\vk, -\vp, \vp-\vk) + C^{(211)}_{ijk}(\vk, -\vp, \vp-\vk)\big)  \nonumber\\
   &+ &   \frac{1}{2} k_ik_jk_k k_l \int \frac{d^3p}{(2\pi)^3} C^{(11)}_{ij}(\vp) C^{(11)}_{kl}(\vk - \vp) \Bigg], 
\end{eqnarray}
where the polyspectra $C^{(n_1 \cdots n_N)}_{i_1\cdots i_N}(\vk_1,\dots,\vk_N)$ at order $(n_1 +\cdots + n_N)$ are defined as
\begin{equation} \label{MatPol}
 \langle \Psi^{(n_1)i_1}(\vk_1)\cdots\Psi^{(n_N) i_N}(\vk_N)\rangle_c = 
 (-i)^{N-2} (2\pi)^3 \delta_D (\vk_1 + \cdots + \vk_N) C^{(n_1 \cdots n_N)}_{i_1\cdots i_N}(\vk_1,\dots,\vk_N).
\end{equation}
The exponential prefactor in Eq.~(\ref{MatPS}) is responsible for the smearing of the BAO features. 
Note also that fourth-order terms, such as $C^{(13)}_{ij}(\vp)$, 
have been neglected in the exponential; if these are considered, the corrections are somewhat large, 
resulting in an over-suppression 
of the BAO peak \cite{Vlah:2014nta}. If we further expand the exponential in Eq.~(\ref{MatPS}) we obtain the 1-loop SPT matter power spectrum \cite{Mat08a}.

We work mainly with the divergence of the displacement field (\ref{deftheta}), which implies $\Psi^i = -i k^i \theta / k^2$. From the polyspectra 
we can write the cumulants for $\theta$ as
\begin{eqnarray}
 \langle \theta^{(m)}(\vk) \theta^{(n)}(\vk') \rangle' &=& k^i k^j C^{(mn)}_{ij}(\vk), \\
 \langle \theta^{(m)}(\vk_1) \theta^{(n)}(\vk_2) \theta^{(p)}(\vk_3) \rangle_c &=&
 - i (2\pi)^3  \delta_D (\vk_1 + \vk_2 + \vk_3) k^i_1 k^j_2 k^k_3 C^{(mnp)}_{i j k}(\vk_1,\vk_2,\vk_3).
\end{eqnarray}
The symbol $\langle \cdots \rangle'$ means that we are omitting a factor $(2\pi)^3 \delta_D(\vk + \vk')$ on the 
two-point cumulant $\langle \cdots \rangle_c$. 
The linear matter power spectrum is simply given by $P_L(k) = \langle \theta^{(1)}(\vk) \theta^{(1)}(\vk') \rangle'$.

In our approach, the $\theta$ functions are modified to
\begin{equation}
 \theta^{(n)} \rightarrow \theta^{(\bar{n})} \equiv \theta^{(n)} + \theta_{\mQ^{(n)}}
\end{equation}
where $\theta_{\mQ^{(n)}} = \sum_{p+q+r+s=n} \theta^{(pqrs)}$. 
Thus, the two-point cumulants are corrected as
\begin{eqnarray}
 \langle \theta^{(2)}(\vk) \theta^{(2)}(\vk') \rangle &\rightarrow& \langle \theta^{(\bar{2})}(\vk) \theta^{(\bar{2})}(\vk') \rangle =
 \langle \theta^{(2)}(\vk) \theta^{(2)}(\vk') \rangle + 2  \langle \theta^{(\bar{2})}(\vk)\theta_{\mQ^{(2)}}(\vk') \rangle,\label{22cumcorr} \\
 \langle \theta^{(3)}(\vk) \theta^{(1)}(\vk') \rangle &\rightarrow&
 \langle \theta^{(\bar{3})}(\vk) \theta^{(\bar{1})}(\vk') \rangle = \langle \theta^{(3)}(\vk) \theta^{(1)}(\vk') \rangle
 +\langle \theta^{(3)}(\vk) \theta_{\mQ^{(1)}}(\vk') \rangle +   \langle  \theta_{\mQ^{(3)}}(\vk)  \theta^{(1)}(\vk')\rangle, \label{31cumcorr}
\end{eqnarray}
where products of the corrections are neglected because they introduce quadratic terms in the velocity dispersion $\sigma_0$.
The matter power spectrum, including the effects of the VDT, is obtained from Eq.~(\ref{MatPS}) by the substitution of unbarred 
orders for barred orders, e.g., $C^{(31)}\rightarrow C^{(\bar{3}\bar{1})}$. 

We further expand the exponential in Eq.~(\ref{MatPS}) to present
results below and in Sec.~\ref{section:LD}.
As an application
we calculate one of the corrections to the power spectrum. By using 
the techniques of this appendix all other contributions can be
calculated. 

To do this, it is useful to consider the following identities: 
given two vectors $\vk$ and $\vp$ and the introduction of the quantities 
\begin{equation}
r \equiv \frac{|\vp|}{|\vk|}, \qquad \qquad  x = \hat{\vk}\cdot\hat{\vp},
\end{equation}
we have
\begin{eqnarray}
 \frac{(\vk-\vp)\cdot\vk}{|\vk-\vp||\vk|} = \frac{1-rx}{\sqrt{1+r^2 - 2 r x}}, \qquad & & \qquad 
 \frac{(\vk-\vp)\cdot\vp}{|\vk-\vp||\vp|} = \frac{x-r}{\sqrt{1+r^2 - 2 r x}}, \\
  \frac{(\vk+\vp)\cdot\vk}{|\vk+\vp||\vk|} = \frac{1+rx}{\sqrt{1+r^2 + 2 r x}}, \qquad & & \qquad 
   \frac{(\vk+\vp)\cdot\vp}{|\vk-\vp||\vp|} = \frac{x+r}{\sqrt{1+r^2 + 2 r x}}. \nonumber
\end{eqnarray}

We are interested in obtaining the corrections that involve the third-order VDT. The only source of 
the Lagrangian displacement equation that contains it is $\mathcal{Q}^{(0003)}$. 
From Eqs.~(\ref{sol3orders}) and (\ref{DefQSources}) we have 
\begin{equation}
 \mathcal{Q}^{(0003)}(\vk,\tau) = k_ik_j \sigma^{(3)ij} = \frac{1}{9} \so f D_+ k_i k_j \mJ^{ij}(\ve k).
\end{equation}
Integrating against the Green function (\ref{Gfunctionsol}) we obtain 
\begin{equation}
\theta^{(0003)}(\ve k,\tau) =  \frac{2}{15} f \frac{ \so }{\h_0^2 }  k_ik_j \sigma^{(3)ij}(\ve k,\tau) D_+^2(\tau)
=  \frac{2}{9\cdot 15} f \frac{\so}{\h_0^2 } k_i k_j \mJ^{ij}(\ve k)  D_+^3(\tau)
\end{equation}

We aim to calculate the cumulant 
\begin{equation}
 \langle \theta^{(1)}(\ve k',\tau) \theta^{(0003)}(\ve k,\tau)\rangle_c =  
    \frac{2}{9\cdot 15} \so  f D_+^4 \frac{1}{\h_0^2 }  k_i k_j  \langle \theta^{(1)}(\ve k')\mJ^{ij}(\ve k) \rangle_c, 
\end{equation}
which is one of several terms that comprise the third term of the rhs of Eq.~(\ref{31cumcorr}).    
To do this we use the first-order, Zel'dovich approximation, identity,
\begin{equation}
\theta^{(1)}(\ve k) = -\delta^{(1)}_0(\ve k),
\end{equation}
and we split the matrix $\mJ^{ij} = \mJ^{ij}_{L^{(1)}}+\mJ^{ij}_{L^{(2)}}+\mJ^{ij}_{L^{(3)}}$, 
depending on which LPT kernel is contained in Eq.~(\ref{app:KK3}).
We separately calculate each contribution:
\begin{eqnarray}
  k_i k_j  \langle \theta^{(1)}(\ve k')\mJ^{ij}_{L^{(1)}}(\ve k) \rangle' &=& 
  - \frac{26}{15}(1+f)(1+2f) \left(\int \frac{d^3p}{(2\pi)^3} P_L(p) \right) k^2 P_L(k),\\
  k_i k_j  \langle \theta^{(1)}(\ve k')\mJ^{ij}_{L^{(2)}}(\ve k) \rangle' &=& 
  - \frac{9}{7}(1+2f) k^2 P_L(k) \times \\ 
  \int  \frac{d^3p}{(2\pi)^3} P_L(p) \!\!\!&\!\!\! &\!\!\! 
 \left( \frac{(\ve k - \ve p)\cdot\ve k}{|\ve k-\ve p| | \ve k|} \frac{(\ve k - \ve p)\cdot\ve p}{|\ve k-\ve p| | \ve p|}
 + \frac{(\ve k + \ve p)\cdot\ve k}{|\ve k+\ve p| | \ve k|} \frac{(\ve k + \ve p)\cdot\ve p}{|\ve k+\ve p| | \ve p|} \right)
  (\hat{\ve k}\cdot\hat{\ve p}-(\hat{\ve k}\cdot\hat{\ve p})^3 ),  \nonumber\\
 k_i k_j  \langle \theta^{(1)}(\ve k')\mJ^{ij}_{L^{(3)}}(\ve k) \rangle' &=& -\frac{20}{21}\left(\int \frac{d^3p}{(2\pi)^3} P_L(p) \right) k^2 P_L(k) \\
+  \frac{5}{7} \!\!\!&\!\!\! &\!\!\!  \left(\int  \frac{d^3p}{(2\pi)^3} P_L(p)
 \left[ \left(\frac{(\ve k - \ve p)\cdot\ve p}{|\ve k-\ve p| | \ve p|} \right)^2 +  \left(\frac{(\ve k + \ve p)\cdot\ve p}{|\ve k+\ve p|  | \ve p|} \right)^2 \right]
 \Big(1-(\hat{\ve k}\cdot\hat{\ve p})^2\Big)\right)  k^2 P_L(k).  \nonumber
\end{eqnarray}
We can sum the three pieces to get
\begin{eqnarray}
 & & k_i k_j  \langle \theta^{(1)}(\ve k')\mJ^{ij}(\ve k) \rangle' = 
  - \frac{1}{5}\left(\frac{94}{7} +\frac{26}{3} f(3+2f) \right) \left(\int \frac{d^3p}{(2\pi)^3} P_L(p) \right)  k^2 P_L(k) \\
& &   -  \frac{9}{7}(1+2f) k^2 P_L(k)  \frac{k^3}{4\pi^2}\int_0^{\infty} dr r^2 P_L(kr) \tilde{R}_{(0003)L_2}(r) 
   +  \frac{5}{7}   k^2 P_L(k)  \frac{k^3}{4\pi^2}\int_0^{\infty} dr r^2 P_L(kr) \tilde{R}_{(0003)L_3}(r).  \nonumber
\end{eqnarray}
The first term on the rhs has contributions from the kernels $K^{(3)}_{L^{(1)}}$ and $K^{(3)}_{L^{(3)}}$, the second from $K^{(3)}_{L^{(2)}}$, 
and the third from $K^{(3)}_{L^{(3)}}$. The functions $\tilde{R}$ are defined as

\begin{eqnarray}
 \tilde{R}_{(0003)L_3}(r) &=& \int_{-1}^1 dx (1-x^2)  \left(\frac{(x-r)^2}{1+r^2 - 2 r x} + \frac{(x+r)^2}{1+r^2 + 2 r x} \right),\\
 \tilde{R}_{(0003)L_2}(r) &=& \int_{-1}^1 dx (1-x^2)x \left(\frac{(1-rx)(x-r)}{1+r^2 - 2 r x} + \frac{(1+rx)(x+r)}{1+r^2 + 2 r x} \right).
\end{eqnarray}

Then, we obtain the correction to the power spectrum due to 
$\langle \theta^{(1)}(\vk) \,\theta^{(0003)}(\vk')\rangle$. This is

\begin{equation}\label{P_1_0003}
P_{1(0003)}(\ve k, \tau) = \frac{2}{135} f D_+^4 \frac{\so}{\h^2_0} k_i k_j \langle \theta^{(1)}(\ve k')\mJ^{ij}(\ve k) \rangle'
\propto - f D_+^4.
\end{equation}
 
\begin{figure}[ht]
	\begin{center}
	\includegraphics[width=4 in]{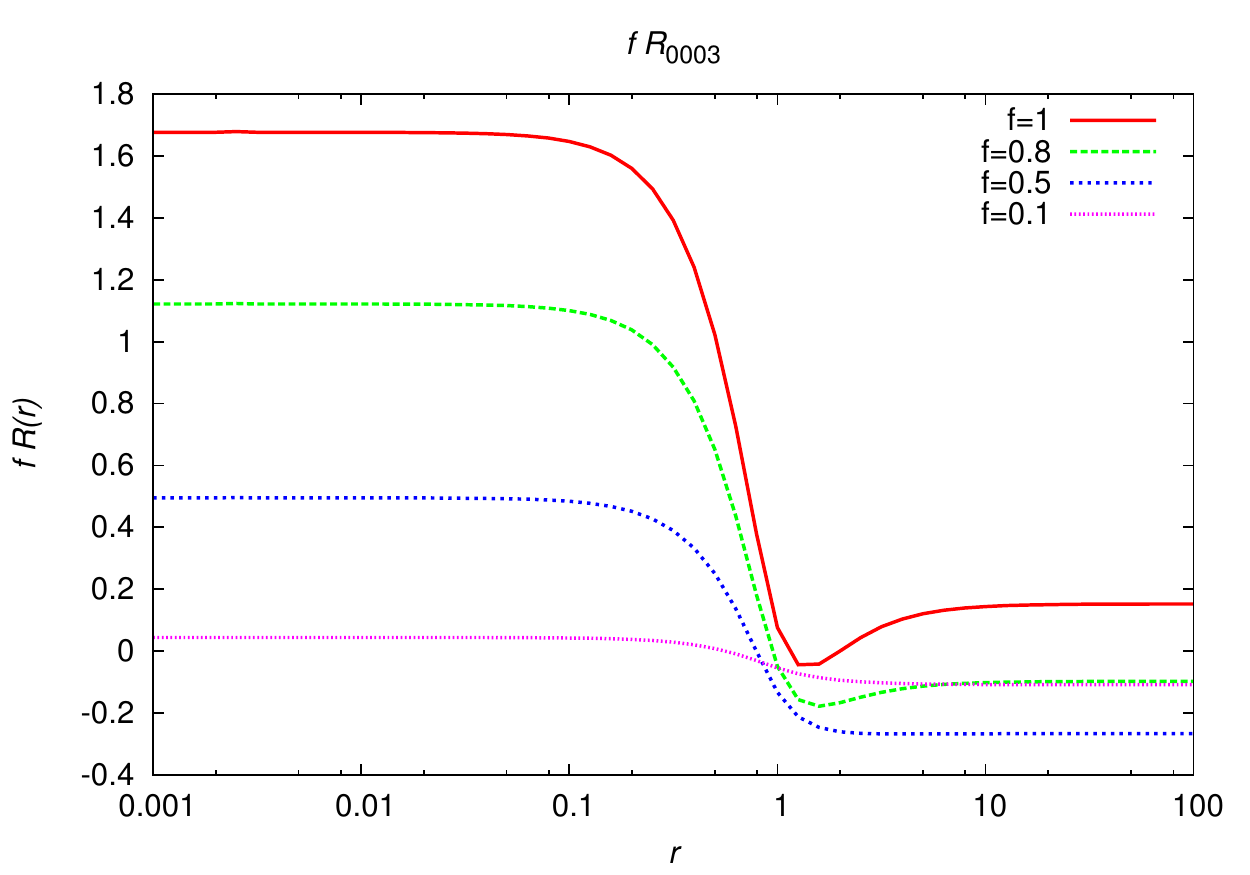}
	\caption{Function $f \times \tilde{R}_{(0003)}(r,f)$ for different values of $f$.
	\label{fig:PlotsR}}
	\end{center}
\end{figure}

It is useful to define the function 
\begin{equation} \label{defineR}
\tilde{R}_{(0003)}(r,f) \equiv \frac{9}{7}(1+2f) \tilde{R}_{(0003)L_2}(r) - \frac{5}{7} \tilde{R}_{(0003)L_3}(r),
\end{equation}
which we plot in  Fig.~\ref{fig:PlotsR} for different values of $f$.

\section{FREE-STREAMING SCALE} \label{app:fss}

An accurate calculation of the free-streaming scale requires a fully relativistic treatment of the linear theory
at early times, starting the integration of Boltzmann-Einstein equations 
well inside the radiation-dominated epoch; see, e.g., Ref.~\cite{Piattella:2015nda}. In our theory, one is tempted to replace 
the linear power spectrum obtained without velocity dispersion by the linear corrections provided by first-order correlations.
In this appendix we do this only to roughly estimate the free-streaming scale of the dark matter and therefore the 
cutoff in the linear power spectrum. To gain insight on the validity of this approach we apply the results to the
case of neutrinos.

We consider the linear correction of the $\theta$ field given in Eq.~(\ref{lcto1}) and take correlations of the linear terms 
to obtain the linear correction to the power spectrum,
\begin{equation} \label{pslc}
 P_L(\vk,\tau) \rightarrow \langle (\theta^{(1)}(\vk)+\theta_{\mathcal{Q}^{(1)}}(\vk) ) (\theta^{(1)}(\vk')+\theta_{\mathcal{Q}^{(1)}}(\vk') )\rangle'
 = \left( 1-\frac{k^2}{k_{\text{fs}}^2}  \right) P_L(\vk,\tau)
\end{equation}
where we defined the comoving free-streaming scale as the cutoff scale in the power spectrum, i.e.,
\begin{equation} \label{FSscale}
 k_{\text{fs}} = \sqrt{\frac{5}{4}\frac{3}{1+2f} \frac{\h^2_0}{\so}}.
\end{equation}

In neutrino physics one is able to calculate the velocity dispersion $\sigma_{\nu,0}$ (usually written $\sigma^2_{\nu,0}$ in neutrino
literature) in terms of its mass by the relation (see, e.g., Ref.~\cite{Lesgourgues:2006nd})
\begin{equation}
 \sigma_{\nu 0} = \frac{15 \zeta(5)}{\zeta(3)} \left(\frac{4}{11}\right)^{2/3} \frac{T_{\gamma 0}^2}{m^2_{\nu}} 
 = \frac{3.65 \times 10^{-7} \,\text{eV}^2}{m^2_{\nu}},
\end{equation}
which is valid in the nonrelativistic regime (for example, for neutrino mass $m_\nu \sim 0.1 \,\text{eV}$, $\sigma_0 \sim 10^{-5}$). 
We use Eq.~(\ref{FSscale}) with $f=1$, obtaining
\begin{equation} \label{fsnuOur}
 k^2_{\nu \,\text{fs}} = \frac{5}{4} \frac{\h^2_0}{\sigma_{\nu,0}} = 0.382 \left(\frac{m_\nu}{1 \text{eV}}\right)^2 \left(\frac{h}{\text{Mpc}}\right)^2.
\end{equation}
Our result is quite similar to that in the neutrino literature: For example, Refs.~\cite{Shoji:2010hm,Ringwald:2004np}
use the mean thermal neutrino velocity $c^2_s$ to arrive at $k^2_{\nu \,\text{fs}} = \frac{3}{2} \frac{\h^2_0}{c^2_s}$ with the identification
of $c^2_s = \sigma_{\nu 0}$; a further correction in \cite{Shoji:2010hm} leads to $c^2_s = 5 \sigma_{\nu 0} / 9$. 
All these results are
obtained in the fluid approximation with a  {\it Jeans-like} mechanism in Euler's 
equation and neglecting the VDT equation, in contrast to our approach.
Similar estimations for $k_\text{fs}$ can be found in the literature, all of 
them in agreement about the order of magnitude. Note, however, that 
different definitions exist for the free-streaming scale.

\section{STOCHASTIC RESIDUAL VECTOR FIELD} \label{app::Gamma}

In this appendix we show that the stochastic residual vector $\Gamma^i$ is transverse and we derive its time dependence.
Let us start by considering the mass conservation relation
\begin{eqnarray} \label{dp1}
 1+\delta(\vx) &=& \int d^3q (1+\delta(\vq)) \delta_D(\vx - \vq - \ve s(\vq,\tau) ) 
 = \int d^3q \Dk k e^{-i\vk\cdot(\vx - \vq - \Psi )} e^{i\vk \cdot \Gamma}(1+\delta(\vq)) \nonumber\\
 &=& \int d^3q \Dk k e^{-i\vk\cdot(\vx - \vq - \Psi )} \Big( 1+ \delta(\vq) e^{i\vk \cdot \Gamma} + (e^{i\vk \cdot \Gamma}-1) \Big)
\end{eqnarray}
The relation (\ref{BE:deltaPsi}) between matter and displacement fields is assumed to hold; thus,
\begin{equation} \label{dp2}
1+\delta(\vx) =  \int d^3q \Dk k e^{-i\vk\cdot(\vx - \vq - \Psi )}.
\end{equation}
From the two previous equations we obtain
\begin{equation} \label{dp3}
 e^{i\vk\cdot \Psi} \Big[ \delta(\vq) e^{i\vk \cdot \Gamma} + (e^{i\vk \cdot \Gamma}-1) \Big]  = 0.
\end{equation}
Note that $\delta(\vq) = \delta(\vx,\tau=\tau_i)$ is a constant field in time. We treat it as a small first-order quantity.
Solving (\ref{dp3}) order by order we have
\begin{equation}
 i \vk \cdot \Gamma^{(0)}  =  0 \quad \Longrightarrow  \quad \Gamma^{(0)} = i \vk \times \Phi^{(0)},
\end{equation}
\begin{equation}
 \Gamma^{(1)}  = i \frac{\vk}{k^2} \delta(\vq) + i \vk \times \Phi^{(1)}, \qquad 
 \Gamma^{(2)}  = -i \frac{\vk}{k^2} \frac{1}{2} (\delta(\vq))^2 + i \vk \times \Phi^{(2)}.
\end{equation}
Stopping at second order is enough to estimate the sources in Eqs.~(\ref{app:sources1order})--(\ref{app:sources3order}).
Note, we can start the evolution early and neglect $\delta(\vq)$, making $\Gamma$ a transverse vector, 
\begin{equation} \label{GeqrotPhi}
 \Gamma = \nabla \times \Phi,
\end{equation}
which is equivalent to the result $J=\bar{J}$ in Sec.~\ref{section:EoM}.
From Eq.~(\ref{GeqrotPhi}) the VDT can be written as
\begin{equation} \label{VDTofGamma}
 \sigma_{ij} = \epsilon_{ikl}\epsilon_{jmn} \langle\nabla^k\dot{\Phi}^l \nabla^m\dot{\Phi}^n \rangle_p.
\end{equation}
From this equation and $\sigma_{ij}^{(n)} \propto D_+^{n-2}$, we can read the time dependence of the $\Gamma$ stochastic field,
\begin{equation}
 \Gamma^{(n)}(\vq,\tau) \propto D_{+}^{\frac{n-1}{2}},  
\end{equation}
while $\Psi^{(n)} \propto D_+^{(n)}$. As a consequence the stochastic source 
$\so^{ik} \nabla_k \Gamma^{(2)l} \nabla_l \dot{\Psi}^{(1)j} \in S^{(3)}_\Gamma (\vq,\tau)$ in Eq.~(\ref{app:sources3order})
grows slower than the source $S^{(3)}_E = \so^{ik} \nabla_k \Psi^{(2)l} \nabla_l \dot{\Psi}^{(1)j}$, and the same holds
for all other sources.  
Moreover, again from Eq.~(\ref{VDTofGamma}), the
$\Gamma$ function has a factor $\sqrt{\sigma_0}$, further suppressing 
the sources as $S^{(n)}_\Gamma \propto \sigma_0^{3/2}$ or even with 
higher powers of $\so$, in contrast to the sources
$S^{(n)}_I$ that are linear in $\sigma_0$. Therefore, if we stick to the prescription of only maintaining leading
orders in $\sigma_0$ in the computations, as we have done so far, all the stochastic sources should be neglected.


\providecommand{\href}[2]{#2}\begingroup\raggedright\endgroup

\end{document}